\documentclass[aps,prl,twocolumn,superscriptaddress,groupedaddress]{revtex4} 
\usepackage{graphicx}  % needed for figures
\usepackage{dcolumn}   % needed for some tables
\usepackage{bm}        % for math
\usepackage{amssymb}   % for math

\usepackage[utf8]{inputenc}
\usepackage[T1]{fontenc}
\usepackage{amsmath}
\usepackage{ae}
\usepackage{color}
\usepackage{graphicx}
\usepackage{bbm}

\usepackage[colorlinks=true, linkcolor=blue, bookmarks=true]{hyperref}
\newcommand{\arXiv}[1]{\href{http://www.arXiv.org/abs/#1}{#1}}

\newcommand{\rd}{\ensuremath{\mathrm{d}}}

\def\be{\begin{equation}}
\def\ee{\end{equation}}

\begin{document}

\title{Gravitational infall in the hard wall model}

\author{Ben Craps} \affiliation{Theoretische Natuurkunde, Vrije Universiteit Brussel, and \\ International Solvay Institutes, Pleinlaan 2, B-1050 Brussels, Belgium}
\affiliation{Laboratoire de Physique Th\'eorique, Ecole Normale Sup\'erieure, 24 rue Lhomond, F-75231 Paris Cedex 05, France}
\author{E.~J.~Lindgren} \affiliation{Theoretische Natuurkunde, Vrije Universiteit Brussel, and \\ International Solvay Institutes, Pleinlaan 2, B-1050 Brussels, Belgium}
\affiliation{Physique Th\'eorique et Math\'ematique, Universit\'e Libre de Bruxelles, Campus Plaine C.P.\ 231, B-1050 Bruxelles, Belgium}
\author{Anastasios Taliotis} \affiliation{Theoretische Natuurkunde, Vrije Universiteit Brussel, and \\ International Solvay Institutes, Pleinlaan 2, B-1050 Brussels, Belgium}
\author{Joris Vanhoof} \affiliation{Theoretische Natuurkunde, Vrije Universiteit Brussel, and \\ International Solvay Institutes, Pleinlaan 2, B-1050 Brussels, Belgium}
\author{Hongbao Zhang} \affiliation{Theoretische Natuurkunde, Vrije Universiteit Brussel, and \\ International Solvay Institutes, Pleinlaan 2, B-1050 Brussels, Belgium}

\date{\today}

\begin{abstract}
 An infalling shell in the hard wall model provides a simple holographic model for energy injection in a confining gauge theory. Depending on its parameters, a scalar shell either collapses into a large black brane, or scatters between the hard wall and the anti-de Sitter boundary. In the scattering regime, we find numerical solutions that keep oscillating for as long as we have followed their evolution, and we provide an analytic argument that shows that a black brane can never be formed. This provides examples of states in infinite-volume field theory that never thermalize. We find that the field theory expectation value of a scalar operator keeps oscillating, with an amplitude that undergoes modulation.    
\end{abstract}

% \pacs{11.25.-w, 11.25.Tq}
\maketitle

{\it Introduction.}
Experimental results on ultrarelativistic heavy ion collisions suggest a fast transition from an initial far-from-equilibrium state to a quark gluon plasma well-described by near-ideal hydrodynamics \cite{Gyulassy:2004zy}. Since small viscosity implies strong coupling, this has motivated the use of gauge/gravity duality to study thermalization (or the approach to hydrodynamics) of certain conformal field theories (CFTs) after a sudden injection of energy \cite{Danielsson:1999zt,Janik:2006gp,Hubeny:2007xt,Kovchegov:2007pq,Lin:2008rw,Chesler:2008hg,Bhattacharyya:2009uu, Beuf:2009cx, Das:2010yw,AbajoArrastia:2010yt,Albash:2010mv,Balasubramanian:2010ce,Heller:2011ju,Heller:2012km,Buchel:2012gw, Balasubramanian:2013rva}. Under the duality, thermalization in the field theory corresponds to black brane formation in the dual bulk theory. An encouraging result is that many such ``holographic'' models indeed give rise to thermalization times that, when extrapolated to real heavy ion collisions, are short enough to 
comfortably accommodate the experimental results \cite{Chesler:2008hg, Beuf:2009cx, AbajoArrastia:2010yt,Albash:2010mv,Balasubramanian:2010ce,Heller:2011ju,Heller:2012km, Balasubramanian:2013rva}. Another remarkable feature is that in the simplest holographic models, the short-wavelength modes thermalize first \cite{Lin:2008rw, Balasubramanian:2010ce}. 
 
In \cite{Craps:2013iaa}, the study of gravitational infall  in the simplest {\em confining} holographic model, namely the hard wall model, was initiated. Perturbative techniques adapted from \cite{Bhattacharyya:2009uu} showed that for sufficiently fast injection of homogeneous energy density, a black brane is formed in the bulk, but that there also exists a regime in which an infalling shell scatters from the hard wall, and then again from the boundary, etc. The intuition is that a black brane is formed if the black brane that would be formed in ordinary anti-de Sitter (AdS) spacetime (without a hard wall) has its event horizon outside the hard wall. (Otherwise the infalling shell is scattered back by the hard wall before it reaches its Schwarzschild radius.) The perturbative techniques used in \cite{Craps:2013iaa} did not allow a reliable study of the long-time evolution of the scattering solutions. Neither did they enable a quantitative study of the transition between both regimes. Another interesting 
question left unanswered in \cite{Craps:2013iaa} is what the scattering solution corresponds to from a field theory perspective. In the present paper, we show that the scattering solutions never collapse, corresponding to field theory states that never thermalize.

Our analysis is related to recent studies, initiated in \cite{Bizon:2011gg}, of whether a spherical shell in anti-de Sitter space will collapse into a small black hole. The picture that has emerged is that depending on the details of the shell, a black hole may be formed (possibly after scattering from the boundary a number of times) or the shell may keep scattering for as long as one can compute the evolution \cite{Maliborski:2013jca, Buchel:2013uba}. This matches well with intricate thermalization behavior of finite volume systems (for solutions that eventually collapse, this was recently discussed in \cite{Abajo-Arrastia:2014fma}). In our work, we are dealing with infinite volume systems, whose thermalization behavior is usually expected to be simpler (see, for instance, \cite{Rigol}).

%%%%%%%%%%%%%%%%%%%%%%%%%%%%%%%%%%%%%%%%%%%%%%%%%%%%%%%%%%%%%%%%%%%%%%%%%%
{\it Holographic setup.}
Our bulk setup is based on Einstein gravity in $d+1$ dimensions with a negative cosmological constant, minimally coupled to a massless scalar field $a$.
The equations of motion are Einstein's equations
\begin{equation}
G_{\mu\nu}-\frac{d(d-1)}{2L^{2}}g_{\mu\nu}-\left(\frac{1}{2}\partial_{\mu}a\partial_{\nu}a-\frac{1}{4}(\partial a)^2g_{\mu\nu}\right)=0
\end{equation}
and the Klein-Gordon equation
\begin{equation}
\frac{1}{\sqrt{-g}}\partial_{\mu}(\sqrt{-g}g^{\mu\nu}\partial_{\nu}a)=0.
\end{equation}
In the dual field theory, we will start with the vacuum state, and inject energy by turning on and off a homogeneous source for the field theory operator dual to the bulk scalar $a=a(z,t)$.  
The corresponding bulk metric ansatz is
\begin{equation} \label{ds}
ds^{2}=\frac{L^{2}}{z^{2}}\left(-f(z,t)e^{-2\delta(z,t)}dt^{2}+\frac{dz^{2}}{f(z,t)}+d\vec{x}^{2}\right),
\end{equation}
where we fix the residual gauge freedom $\delta(z,t)\mapsto\delta(z,t)+p(t)$  by requiring the UV boundary condition $\lim_{z\rightarrow0}\delta(z,t)=0$. At early times, we start from the AdS metric, $f=1$ and $a=\delta=0$. The field theory source is given by the boundary profile of $a$, which we choose to be Gaussian,
\begin{equation}
a_{0}(t)\equiv a(z=0,t)=\epsilon\,e^{-\frac{t^2}{\delta t^2}}. \label{gaussian}
\end{equation} 
Writing prime and dot for differentiation with respect to $z$ and $t$, respectively, and introducing $A\equiv a'$ and $\Pi \equiv e^{\delta}\dot{a}/f$, the equations of motion reduce to
\begin{subequations}\label{eom}
\begin{align}
\dot{A}=&\left(fe^{-\delta}\Pi\right)',\label{Ad} \\
\dot{\Pi}=&z^{d-1}\left(\frac{fe^{-\delta}A}{z^{d-1}}\right)',\label{Pid}\\
\dot{f}=&\frac{z}{d-1}f^{2}e^{-\delta}A\Pi, \label{fd}\\
f'=&\frac{z}{2(d-1)}f\left(A^{2}+\Pi^{2}\right)+\frac{d}{z}(f-1),\label{fp}\\
\delta'=&\frac{z}{2(d-1)}\left(A^{2}+\Pi^{2}\right)\label{dp}.
\end{align}
\end{subequations}

A hard wall is introduced by restricting the range of the $z$ coordinate to $0<z<z_0$, where the location of the hard wall is inversely proportional to the confinement scale $\Lambda$ of the boundary theory, $\Lambda \sim 1/z_0$. At the hard wall, we will mainly consider two possible boundary conditions on the scalar field: Dirichlet boundary conditions 
$0=\left.a\right|_{z=z_{0}}$, corresponding to $\Pi=0$, or Neumann boundary conditions $0=\left.a'\right|_{z=z_{0}}$, corresponding to $A=0$. For the numerical analysis we use the rescaling freedom of the coordinates $(z,t,\vec{x})$ to set $z_{0}=1$. Therefore time $t$ and the injection time $\delta t$ appearing in the plots are given in units of $z_{0}$. The amplitude $\epsilon$ is a dimensionless quantity.

%%%%%%%%%%%%%%%%%%%%%%%%%%%%%%%%%%%%%%%%%%%%%
{\it  Numerical solution.}
 To solve the system numerically, we discretized the equations in the bulk coordinate $z$ using a pseudospectral method based on Chebychev polynomials, see \cite{Wu:2012rib}. In contrast to \cite{Wu:2012rib} where a small cutoff close to the boundary is used to avoid the singularity in eq \eqref{Pid}, we have redefined $A(z,t)=z\tilde{A}(z,t)$. This also proves to be crucial for stable long time evolutions in the $d=4$ case.

\begin{figure}[t]
\centering \label{VEVs}
\includegraphics[scale=0.45]{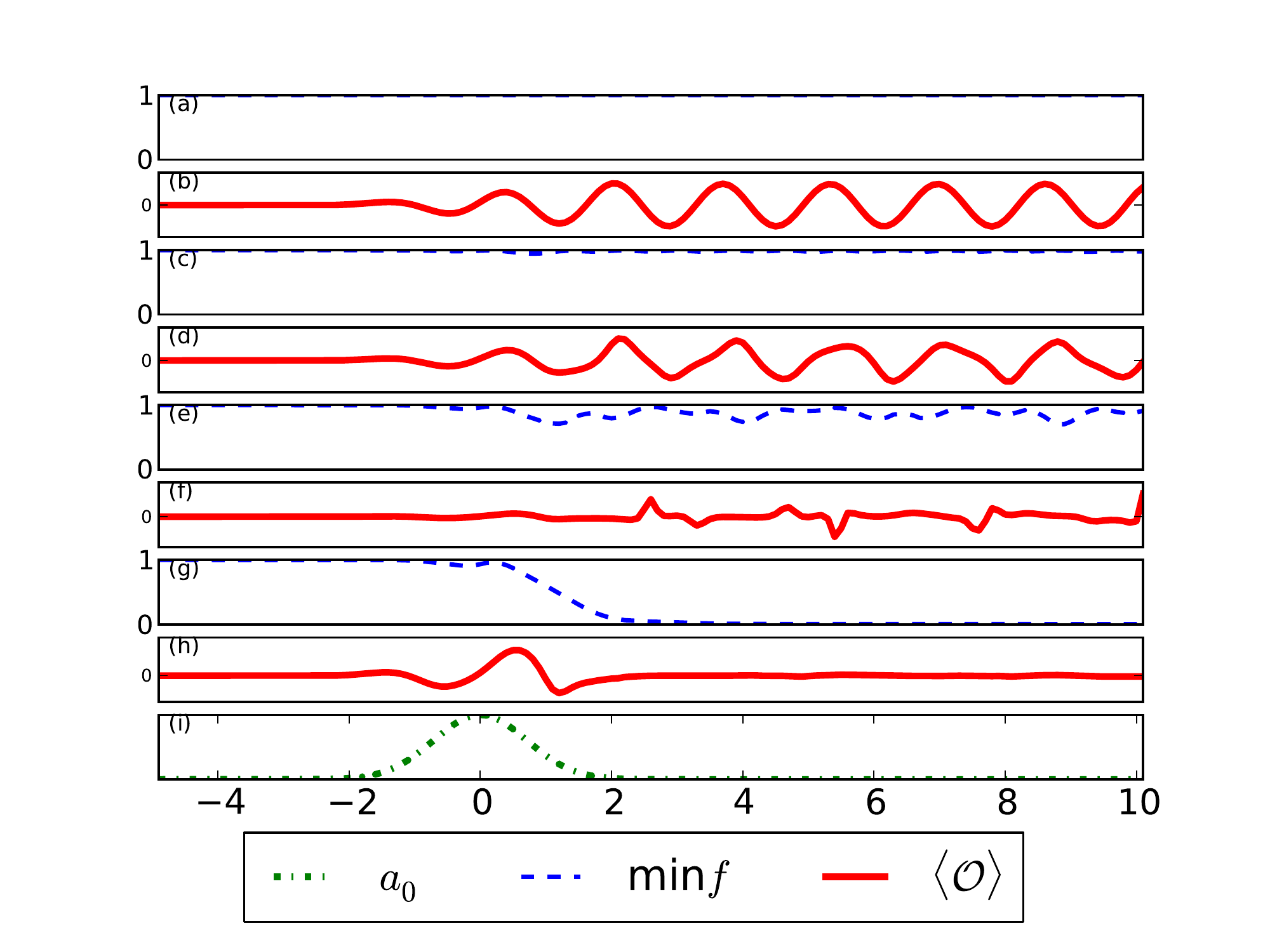}
\caption{\label{evol} Time evolution of various quantities for $d=4$, fixed $\delta t=1$ and Neumann boundary conditions. Panels (a,c,e,g) show the evolution of the minimum of $f$, for three scattering solutions ($\epsilon=0.1$, $\epsilon=0.6$ and $\epsilon=1$) and black brane formation ($\epsilon=1.15$), respectively. Panels (b,d,f,h) contain the corresponding time evolution of the expectation values of the scalar operator in the dual field theory. The last panel shows the profile of the scalar source.}
\end{figure}

Black hole formation is signalled by the formation of an apparent horizon, where the blackening factor $f$ vanishes. Since in the coordinate system (\ref{ds}) an apparent horizon is only reached at infinite time, in practice we declare that a black brane has been formed whenever the minimum of $f$ goes below a cutoff we choose to be $0.02$. In panels (a,c,e,g) of Figure \ref{evol}, we illustrate the evolution of the minimum of the metric function $f$ for $d=4$ field theory dimensions, Neumann boundary conditions, $\delta t=1$ and several values of $\epsilon$.

\begin{figure}[t] \label{phasedi}
\centering 
\includegraphics[scale=0.45]{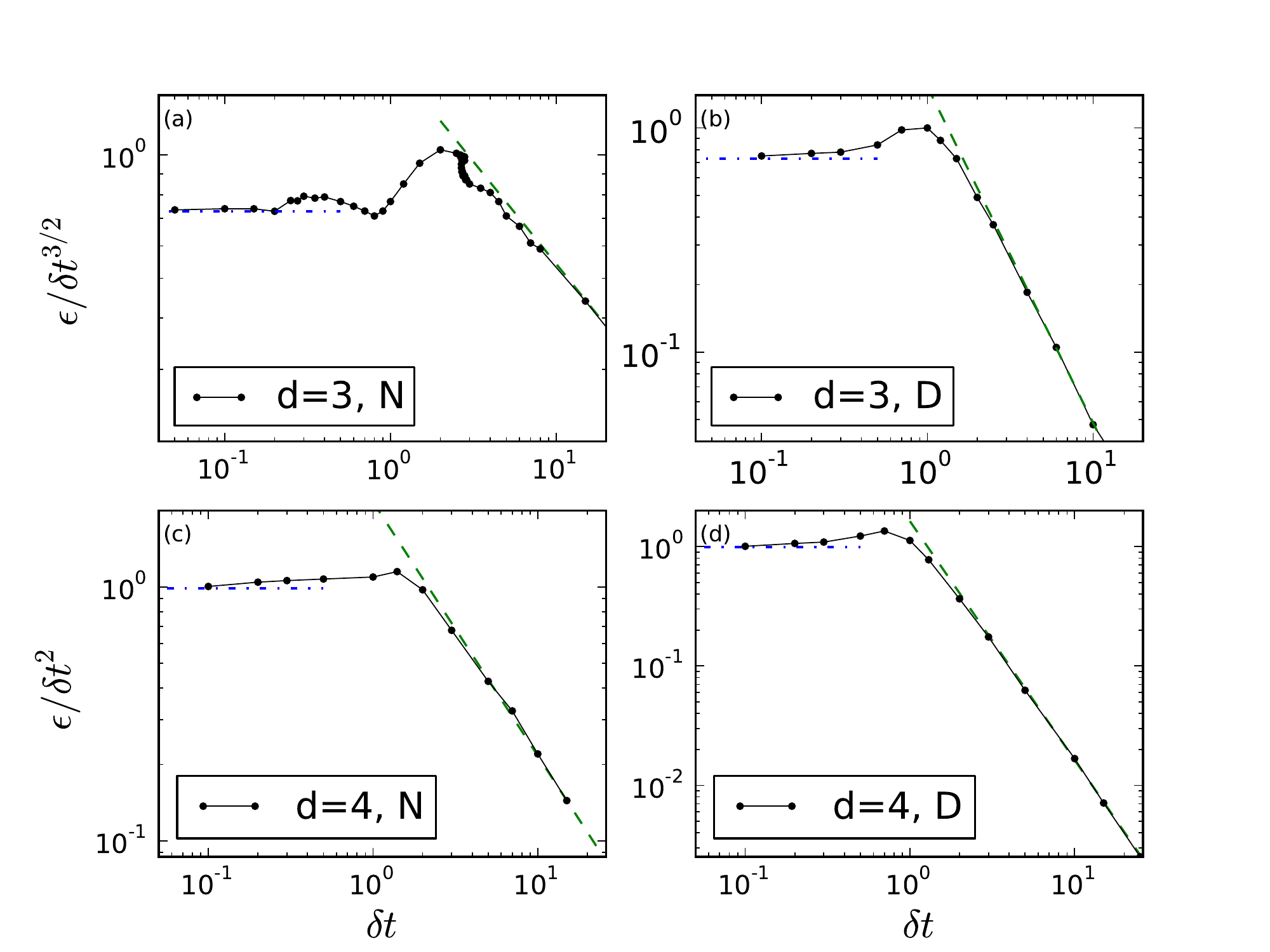}
\caption{\label{eps_delta}Dynamical phase diagrams showing the critical parameters that separate the scattering regime from the black brane regime. The different panels show different dimensions and different boundary conditions, with panels (a,b,c,d) representing d=3 Neumann, d=3 Dirichlet, d=4 Neumann and d=4 Dirichlet, respectively. The dots are the numerically computed critical parameters for gravitational collapse. Above (below) the dots we have the black brane (scattering) phase. The straight lines correspond to analytically computed asymptotic behaviors, as explained in the text.}
\end{figure}

Figure \ref{eps_delta} shows the dynamical phase diagrams for $d=3$ and $d=4$, indicating in which parameter regions a black brane is formed, and in which a scattering solution is found.

%%%%%%%%%%%%%%%%%%%%%%%%%%%%%%%%%%%%%%%%%%%%%%%%%%%%%%%%%%%%%%%%%%%%%%%%%%%%%%%%%%%%%
{\it Asymptotic behavior of the dynamical phase diagram.} To have a better understanding of our numerical results, we now investigate analytically the regimes of very small and very large injection time  $\delta t$. In the regime $\Lambda\delta t \ll1$, it is expected that a black brane will form  if the black brane formed in pure AdS would have its horizon outside the location of the hard wall \cite{Craps:2013iaa}.  For $d=3$ and  small $\epsilon$, the mass density of the black brane is given by $M=(1/2)\int_{-\infty}^{\infty}\text{d}t\left(\ddot{a_0}(t)\right)^{2}$ \cite{Bhattacharyya:2009uu}. The critical parameters correspond to $M=1/z_0^3$, which for the profile \eqref{gaussian} gives $\epsilon_c=\left(8/(9\pi)\right)^{1/4}\delta t^{3/2}z_0^{-3/2}$. For $d=4$, the mass density of the black brane is given by formula (B.14) in \cite{Bhattacharyya:2009uu}, which for the profile \eqref{gaussian} results in the critical parameter $\epsilon_c=\left(3/\pi\right)^{1/2}\delta t^{2}z_0^{-2}$. 

The regime $\Lambda\delta t \gg1$ corresponds to adiabatic energy injection. For Neumann boundary conditions, $A=0$ at the hard wall and the assumption of slow injection will be $\dot{\Pi}=\dot{A}=\dot{f}=0$. From equation \eqref{Pid} we then find that $A=Cz^{d-1}f^{-1}e^{\delta}$, where $C$ is an integration constant, and using the Neumann boundary condition we must have $C=0$ and so $A=0$ throughout the bulk. Assuming the boundary conditions $\Pi(z=0)=\lambda$, \eqref{Ad} now gives $\Pi=f^{-1}e^{\delta}\lambda$ since $f(z=0)=1$ and $\delta(z=0)=0$. Using these solutions in \eqref{fp} and \eqref{dp}, we obtain an ordinary differential equation (ODE), conveniently written in terms of $S\equiv fe^{-\delta}$, as
\begin{equation}
 S'=\frac{d}{z}\left(S-e^{-\frac{\lambda^2}{2(d-1)}\int_0^z z'S(z')^{-2} \rd z' } \right),\label{ode}
\end{equation}
where we have used the boundary condition $\delta(z=0)=0$. This ODE can be solved numerically. However, it turns out there is a critical $\lambda_c$ such that for $|\lambda|\geq\lambda_c$ this ODE is not solvable anymore and this indicates that we have a black brane solution instead. For $d=3$ we have $\lambda_c\approx 1.47$ and for $d=4$ we have $\lambda_c\approx 1.85$. Going back to our original setup, $\lambda$ will be time-dependent and equal to the time derivative of the boundary condition of the scalar field. Thus we draw the conclusion that for large injection times, a black hole is formed if we have $\textrm{max}\{|\Pi(z=0,t)|\}\geq \lambda_c$. For the profile given in equation \eqref{gaussian}, we obtain the relation $\epsilon_c=\lambda_c \sqrt{e}\delta t/\sqrt{2}$.

For Dirichlet boundary conditions, we approximate the boundary condition $a_0$ as time-independent. We can then solve \eqref{Pid} to obtain $A=\alpha z^{d-1}S^{-1}$, where $\alpha$ is a constant, related to $a_0$ by the requirement that $a=a_0+\int_0^zA(x) \rd x$ should vanish on the hard wall. Following the same argument as for the Neumann boundary condition, we obtain 
\begin{equation}
 S'=\frac{d}{z}\left(S-e^{-\frac{\alpha^2}{2(d-1)}\int_0^z S(z')^{-2} z'^{2d-1}\rd z' } \right)\label{ode2}.
\end{equation}
We find critical parameters, $a_{0,c}\approx 1.53$ for $d=3$ and $a_{0,c}\approx 1.63$ for $d=4$, beyond which no solution exists. For the profile \eqref{gaussian}, this leads to the critical amplitude $\epsilon_c=a_{0,c}$. As shown in Figure~\ref{eps_delta}, our numerical results are in excellent agreement with the various asymptotic regimes we have explored here.

{\it Weakly non-linear perturbation theory.}
In the case of global AdS, the instability discovered in \cite{Bizon:2011gg} was accompanied by weakly turbulent behaviour due to resonances in the spectrum of linear perturbations (see also \cite{Dias:2012tq, Maliborski:2013jca, Maliborski:2014rma, Balasubramanian:2014cja}). In our setting, we have checked that the frequencies of linearized modes generically do not display obvious resonances. For example for Dirichlet boundary conditions they are given by $\omega_{n}=\gamma^{(d/2)}_{n}/z_{0}$, where $\gamma^{(\nu)}_{n}$ is the $n^{\text{th}}$ zero of $J_{\nu}(x)$; while this spectrum is asymptotically resonant, it is not obviously resonant. The only exception is AdS$_{4}$ with Neumann boundary conditions, where the spectrum of frequencies $\omega_{n}=\pi n/z_{0}$ is resonant. This implies the presence of secular terms in the perturbation analysis. See the appendix for more details.

%%%%%%%%%%%%%%%%%%%%%%%%%%%%%%%%%%%%%%%%%%%%%%%%%%%%%%%%%%%%%%%%%%%%%%%%%%%%%%%%%%%%%%%
{\it Field theory interpretation of the scattering solution.}
It is well-known that black brane formation corresponds to thermalization in the dual field theory. Here we investigate what the scattering solutions we have found correspond to. Holographic renormalization relates expectation values of gauge-invariant operators to the asymptotic behavior of the corresponding bulk fields. We summarize the key results for $d=3$ and $d=4$. See the appendix for details on the (standard) computations (see, for instance, \cite{Skenderis:2002wp, Papadimitriou:2011qb}).

For $d=3$, $\langle\mathcal{O}\rangle$ is given in terms of the near-boundary expansion $a(z,t)=a^{(0)}(t)+a^{(1)}(t)z+a^{(2)}(t)z^2+a^{(3)}(t)z^3+...$ by $\langle\mathcal{O}\rangle =3a^{(3)}$. For the perturbative small-$\epsilon$ scattering solution of \cite{Craps:2013iaa} with Neumann boundary conditions, this yields
\begin{equation} \label{oscO}
\langle\mathcal{O}(t)\rangle=\dddot{a}_{0}(t)+2\sum_{m=1}^{\infty}\dddot{a}_{0}(t-2mz_{0})+\mathcal{O}\left(\epsilon^{3}\right),
\end{equation}
corresponding to an oscillating behavior as a function of time. For $d=4$, we find $\langle{\cal O}\rangle=4a^{(4)}(t)$ (up to a scheme-dependent contribution that vanishes when the source vanishes), yielding a similar oscillating behavior for small $\epsilon$. Our numerical result displayed in panel (b) in Figure~\ref{evol} confirms this behavior. Panels (d,f) show less regular behavior closer to the black brane formation regime, while panel (h) shows that $\langle\mathcal{O}\rangle$ vanishes after a black brane has been formed.

For $d=3$, $\langle {T_{\mu \nu}}\rangle$ is given by $\langle T_{tt}\rangle=2\langle T_{x_{i}x_{i}}\rangle=-f^{(3)}(t)$, where $f^{(3)}$ is the $z^3$-coefficient of the near boundary expansion of $f(z,t)$. To lowest nontrivial order in $\epsilon$, we find using the result of \cite{Craps:2013iaa} for Neumann boundary condition that
\begin{align}
\langle T_{tt}\rangle=&-\frac{1}{2}\int_{0}^{t}\text{d}\tau\Big(\dot{a}_{0}(\tau)\dddot{a}_{0}(\tau) \nonumber \\
&+2\sum_{m=1}^{\infty}\dot{a}_{0}(\tau)\dddot{a}_{0}(\tau-2mz_{0})\Big)+\mathcal{O}\left(\epsilon^{4}\right). \label{Ttt}
\end{align}
Also numerically, we have found that the energy density $\langle T_{tt}\rangle$ (and therefore the pressure $\langle T_{x_{i}x_{i}}\rangle$) reaches a constant value after the source has been turned off. For $d=4$, the results are very similar, except that there is a conformal anomaly in the time range where the source is non-vanishing.

\begin{figure}[t]
\centering 
\includegraphics[scale=0.45]{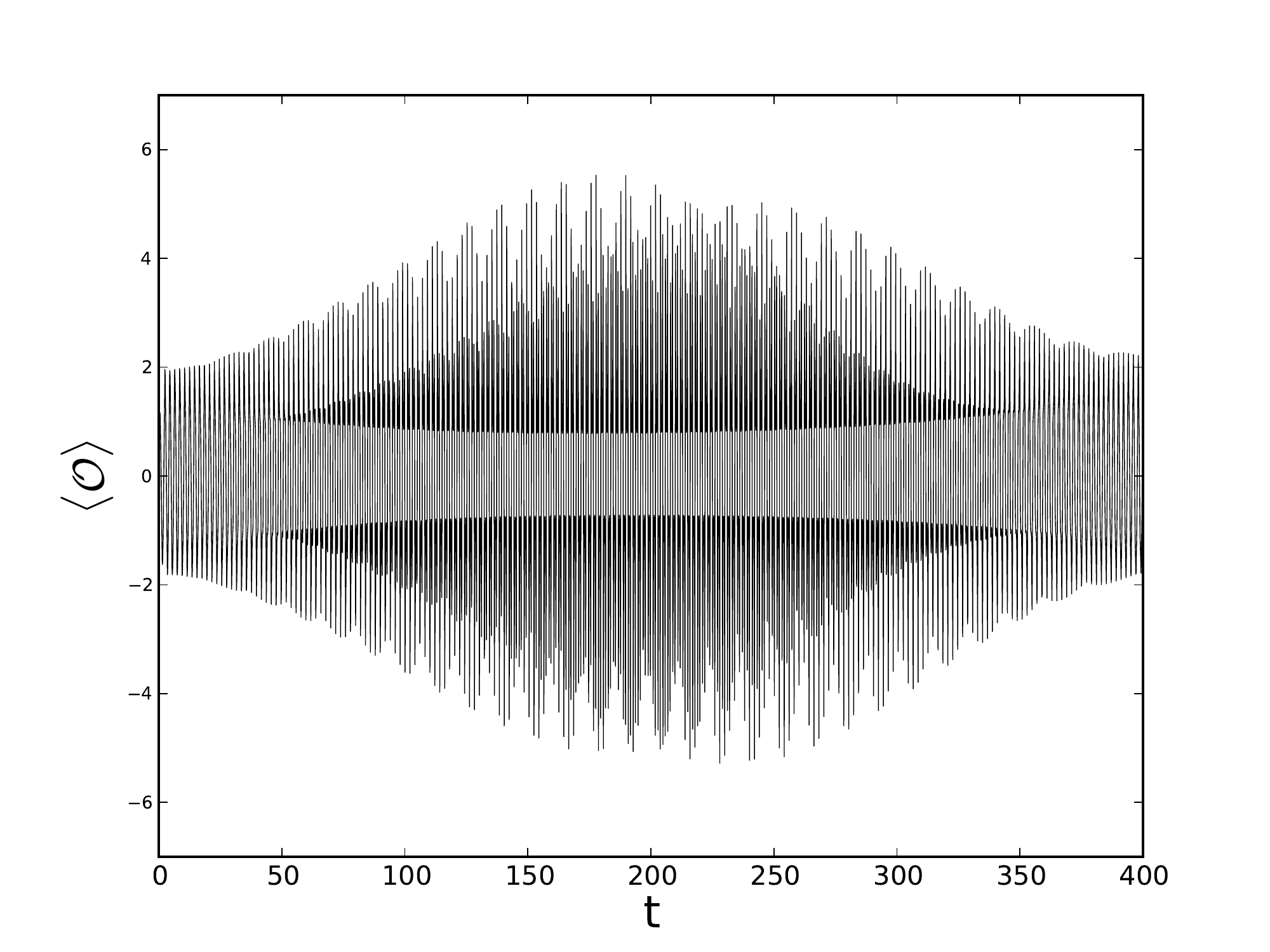}
\caption{\label{longtime} Long-time evolution of $\langle\mathcal{O}\rangle$ in the scattering phase for $d=3$, $\delta t=1$, $\epsilon=0.2$ and Neumann boundary conditions. }
\end{figure}

%%%%%%%%%%%%%%%%%%%%%%%%%%%%%%%%%%%%%%%%%%%%%%%%%%%
{\it Late-time behavior.}
The most noteworthy feature of our results is the oscillating behavior of $\langle\mathcal{O}\rangle$ in the scattering phase. If this behavior persists for all times, it indicates that the out-of-equilibrium state created by the energy injection never thermalizes. While analogous solutions have been found before for field theories in finite volume (dual to asymptotically global AdS spacetimes), to our knowledge this would be the first such example in infinite volume. We have therefore investigated the behavior of $\langle\mathcal{O}\rangle$ in our scattering solutions for much later times than those displayed in Figure~\ref{evol}. Figure~\ref{longtime} shows, for $d=3$ and Neumann boundary conditions, that the oscillations continue for as long as we have followed the evolution, but that they are modulated on a larger timescale. Preliminary results indicate that this timescale decreases with $\epsilon$, roughly like $1/\epsilon^{2}$. We expect that this scaling is due to the above-mentioned secular terms in weakly non-linear perturbation theory. For other dimensions and for other boundary conditions, for which there are no secular terms in perturbation theory, we find a less pronounced modulation.

In fact, a simple analytic argument shows that the scattering solutions can never evolve into a black brane solution. First, since they do not have enough energy to form a large black brane (with horizon in the physical part of spacetime), the only possibility would be a small black brane (with would-be horizon behind the hard wall). However, for Dirichlet or Neumann boundary conditions, (\ref{fd}) implies that $f$ is constant at the hard wall, so $f=1$ if we start from empty AdS. Since small black brane solutions have $f<1$ at the hard wall, they cannot be formed. This conclusion would still hold if at the hard wall we allowed more general boundary conditions $n^{\mu}\partial_{\mu}a=F(a)$, i.e. $z_{0}\sqrt{f}a'=F(a)$, where $F$ is an arbitrary function. These boundary conditions can be imposed in agreement with the variational principle if we add to the bulk action a boundary term at the hard wall proportional to $\mathcal{S}_{b}\sim\int_{z=z_{0}}(\int_{0}^{a}F(b)\text{d}b)\sqrt{\gamma}\text{d}^{d}\mathbf{x}
$. In that case (\ref{fd}) implies that at the hard wall location, $\sqrt{f}=(\int_{0}^{a}F(b)\text{d}b)/(2d-2)+C$, with constant $C$. Again, if initially we have $f=1$ and $a=0$, then at late times we cannot have $f<1$ and $a=0$, as would have been the case for a small black brane. While for these more general boundary conditions we cannot exclude that the system might approach another static solution than a black brane, we have seen no hints of this in our numerical solutions. 

\begin{figure}[t]
\centering 
\includegraphics[scale=0.45]{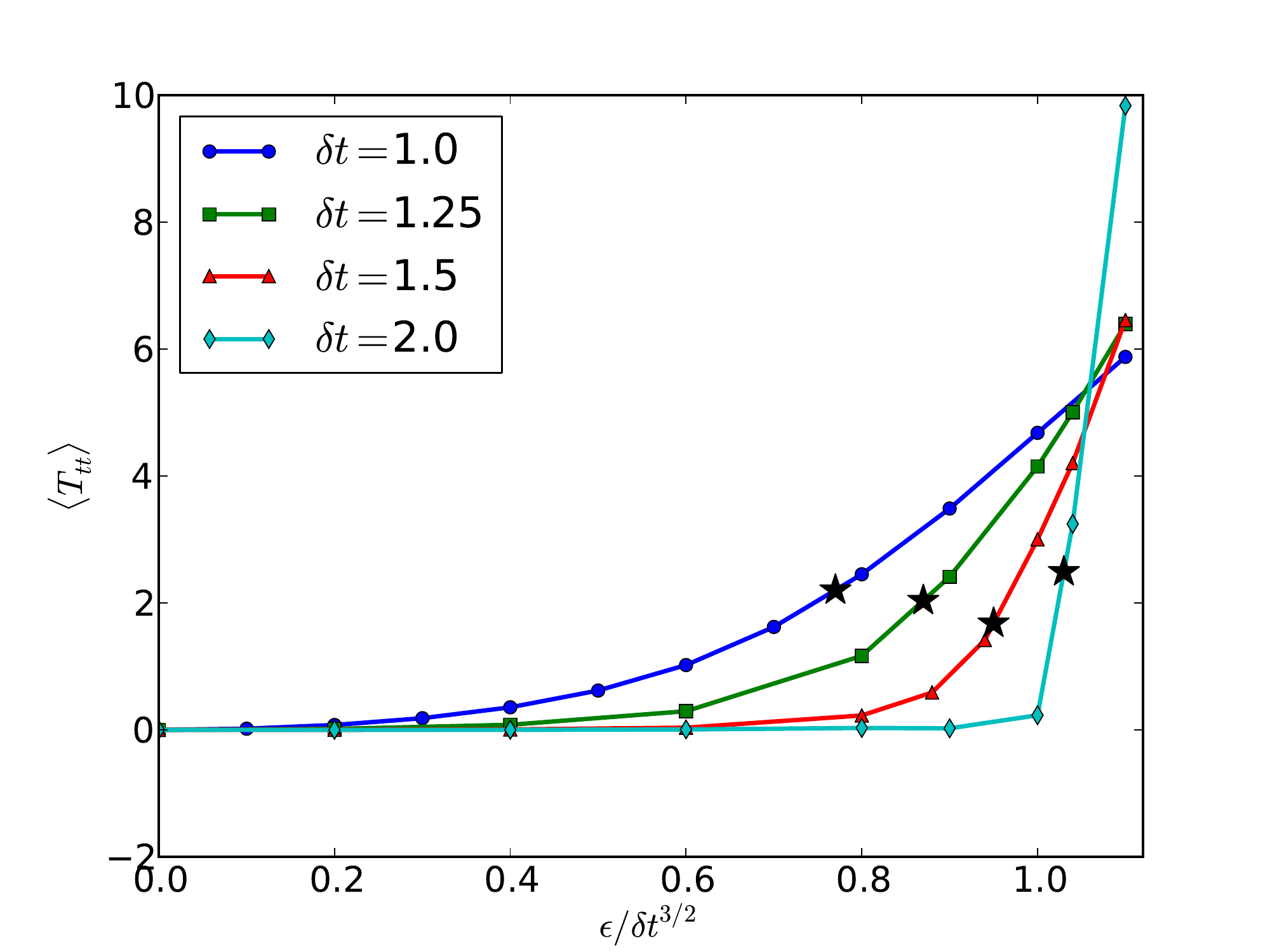}
\caption{\label{T_eps} Total injected energy as a function of $\epsilon$ for $d=3$ and Neumann boundary conditions. The black stars mark the critical value for black hole formation. The scattering regime is on the left of these markers and the black brane formation regime is on the right.}
\end{figure}

Another interesting question is what is the energy density of the final state as a function of the source amplitude $\epsilon$ and the injection time $\delta t$. Figure~\ref{T_eps} shows that for fixed $\delta t$, the injected energy increases as a function of $\epsilon$. While for small $\delta t$ the increase is gradual, for large $\delta t$ the injected energy density is very small in the scattering phase (as can be expected for a source that is turned on and off almost adiabatically), but increases very sharply when the threshold for black brane formation is crossed. (In the latter regime, we have discussed before that the adiabatic approximation breaks down.)

{\it Possible implications for QCD.}
While the hard wall model is only a crude model for QCD, it is tempting to speculate on what our results might mean when extrapolated to real QCD. For instance, if we took the scalar field $a$ to be the dilaton, the source $a_0$ would couple to the operator ${\cal O}={\rm Tr}\,F_{\mu\nu}F^{\mu\nu}\sim {\rm Tr}(E^2-B^2)$. For sufficiently fast injection, the system thermalizes into a deconfined plasma, and $\langle{\rm Tr}(E^2-B^2)\rangle=0$ (as could be expected based on equipartition between electric and magnetic gluon polarizations \footnote{We thank Y.~Kovchegov for correspondence on this.}). For sufficiently slow injection, the degrees of freedom remain confined. The oscillations in ${\rm Tr}\,F_{\mu\nu}F^{\mu\nu}$ can be interpreted in terms of convertions of a collection of glueballs into another collection (and back) \footnote{We thank N.~Evans for suggesting this.}.

\vspace{0.05in}

{\it Acknowledgements.}
We would like to thank Z.~Cao, W.~De Roeck, N.~Evans, O.~Evnin, U.~G\"ursoy, J.~Kapusta, E.~Kiritsis, Y.~Kovchegov, F.~Larsen, L.~McLerran, I.~Papadimitriou, C.~Rosen, Y.~Tian and the organizers and participants of the STAG workshop on holography, gauge theory and black holes for useful discussions. In addition, we thank Y.~Kovchegov for comments on the manuscript. This work was supported in part by the Belgian Federal Science Policy Office through the Interuniversity Attraction Pole P7/37, by FWO-Vlaanderen through project G020714N,  and by the Vrije Universiteit Brussel through the Strategic Research Program ``High-Energy Physics''. The work of EJL was partially supported by the ERC Advanced Grant ``SyDuGraM", by IISN-Belgium (convention 4.4514.08) and by the ``Communaut\'e Fran\c{c}aise de Belgique" through the ARC program. The work of AT is funded by the VUB Research Council. JV is Aspirant FWO.

\appendix

\section{Appendix}

{\it Weakly non-linear perturbations and instabilities.}
Following the approach of \cite{Bizon:2011gg}, we study the possible presence of weakly turbulent instabilities. Starting from initial data $(a(z,t)|_{t=0},\dot{a}(z,t)|_{t=0})=(\epsilon f(z),\epsilon g(z))$ of order $\epsilon$ and using the derived equations of motion, we search for a perturbative solution in $\epsilon$
\begin{equation}\label{pertexp}
a=\sum_{k=0}^{\infty}\epsilon^{2k+1}a_{2k+1}\,,\,\,
\delta=\sum_{k=1}^{\infty}\epsilon^{2k}\delta_{2k}\,,\,\,
f=1+\sum_{k=1}^{\infty}\epsilon^{2k}f_{2k}\,.
\end{equation}
At first order in the $\epsilon$-expansion, we find the linear homogeneous differential equation
\begin{equation}
\ddot{a}_{1}+La_{1}=0, \,\,
L\equiv-z^{d-1}\partial_{z}\left(z^{1-d}\partial_{z}\right).
\end{equation}
The Sturm-Liouville operator $L$ is self-adjoint on the subspace of $L^{2}\left([0,z_{0}],z^{1-d}\text{d}z\right)$ of functions $\psi(z)$ that satisfy the boundary conditions $0=\left.\psi\right|_{z=0}=\left.\alpha\psi+\beta z\partial_{z}\psi\right|_{z=z_{0}}$ for arbitrary real constants $\alpha$ and $\beta$. The inner product on this Hilbert space is $\langle\psi,\chi\rangle\equiv\int_{0}^{z_{0}}\bar{\psi}(z)\chi(z)z^{1-d}\text{d}z$.

$\bullet$ Restricting to the subspace of functions that satisfy Dirichlet boundary conditions $0=\left.\psi\right|_{z=z_{0}}$ at $z=z_0$, we find the orthonormal basis of eigenfunctions of $L$
\begin{equation}
e_{n}(z)=k_{n}^{-1}z^{\frac{d}{2}}J_{\frac{d}{2}}\left(\gamma^{(\frac{d}{2})}_{n}z/z_{0}\right),\,\, n=1,2,...\, ,
\end{equation}
where $k_{n}=z_{0}\,J_{d/2+1}\left(\gamma^{(d/2)}_{n}\right)/\sqrt{2}$ is a normalisation constant that ensures orthonormality $\langle e_{n},e_{m}\rangle=\delta_{nm}$, and where $J_{\nu}(x)$ is the Bessel function of order $\nu$. The eigenfunctions $e_{n}(z)$ correspond to the eigenvalues $\omega_{n}^{2}=(\gamma^{\left(d/2\right)}_{n}/z_{0})^{2}$, where $\gamma^{(\nu)}_{n}$ is the $n$-th zero of $J_{\nu}(x)$, such that $Le_{n}(z)=\omega_{n}^{2}e_{n}(z)$.

$\bullet$ Similarly, restricting to the subspace of functions that satisfy mixed boundary conditions $0=\left.\alpha\psi+z\partial_{z}\psi\right|_{z=z_{0}}$, one finds the orthonormal basis of eigenfunctions,
\begin{equation}
\tilde{e}_{n}(z)=\tilde{k}_{n}^{-1}z^{\frac{d}{2}}J_{\frac{d}{2}}\left(\tau^{(\frac{d}{2},\alpha)}_{n}z/z_{0}\right),\,\, n=1,2,... \,\,
\end{equation}
with normalisation constant
\begin{equation}
\tilde{k}_{n}=\frac{z_{0}}{\sqrt{2}}\,J_{\frac{d}{2}}\left(\tau^{(\frac{d}{2},\alpha)}_{n}\right)\left(1+\frac{\alpha(\alpha+d)}{(\tau^{(\frac{d}{2},\alpha)}_{n})^{2}}\right)^{\frac{1}{2}}.
\end{equation}
The corresponding eigenvalues are $\tilde{\omega}_{n}^{2}=(\tau^{(d/2,\alpha)}_{n}/z_{0})^{2}$ where $\tau^{(\nu,\alpha)}_{n}$ is the $n$-th zero of the function defined as $f(x)=(\alpha+\nu)J_{\nu}(x)+xJ'_{\nu}(x)$.

For global AdS, there are countably many distinct frequencies $(\omega_{i},\omega_{j},\omega_{k},\omega_{l})$ that satisfy the resonance condition $\omega_{l}=\omega_{i}+\omega_{j}-\omega_{k}$ \cite{Bizon:2011gg}. Due to the transcendental behaviour of the Bessel functions, generically there can be no such resonances in our setup. Intuitively, such a fine-tuning of the frequencies can be thought as a consequence of the highly symmetric behavior of the AdS space, which, in our case, is spoilt by the presence of the IR cut-off of the geometry at $z=z_{0}$.

The only notable exception in this regard is AdS$_{4}$ with Neumann boundary conditions at the hard wall. Using the identity $(3/2)J_{3/2}(x)+xJ'_{3/2}(x)=xJ_{1/2}(x)$, one finds that in this case the frequencies of the modes are given by $\omega_{n}=\gamma^{(1/2)}_{n}/z_{0}=\pi n/z_{0}$, which obviously results in a resonant spectrum.

%%%%%%%%%%%%%%%%%%%%%%%%%%%%%
{\it  Holographic renormalization.}
The stress-energy tensor $\langle T_{ij}\rangle$ and scalar $\langle{\cal O}\rangle$ expectation values can be extracted by applying the standard techniques of holographic renormalisation \cite{Skenderis:2002wp}. One has to evaluate the bulk action
\begin{align}
\mathcal{S}=&\int_{\mathcal{M}}\text{d}^{d+1}\mathbf{x}\sqrt{-g}\left(R+d(d-1)-\frac{1}{2}(\partial a)^2\right) \nonumber \\
&+\int_{\partial\mathcal{M}}\text{d}^{d}\mathbf{x}\sqrt{-\gamma}\,2K+\mathcal{S}_{C}
\end{align}
for the on-shell field solutions and then determine the variations $\langle T_{ij}\rangle=-z^{2}(\delta S/\delta\gamma^{ij})$ and $\langle{\cal O}\rangle=\delta S/\delta\phi$.

$\bullet$ In $d=3$, the counterterms are given by
\begin{equation}
\mathcal{S}_{C}=\int_{\partial\mathcal{M}}\text{d}^{3}\mathbf{x}\sqrt{-\gamma}\left(-4-\hat{R}+\frac{1}{2}(\hat{\partial}a)^{2}\right).
\end{equation}
We can read off the expectation values $\langle{\cal O}\rangle=3a^{(3)}(t)$ and $\langle T_{tt}\rangle=2\langle T_{x_{i}x_{i}}\rangle=-f^{(3)}(t)$, where $a^{(3)}(t)$ and $f^{(3)}(t)$ are the $z^{3}$-coefficients in the near boundary expansion, $a(z,t)=a_{0}(t)+za^{(1)}(t)+z^{2}a^{(2)}(t)+z^{3}a^{(3)}(t)+...\,$. The trace of the stress-energy tensor is identically zero, $\langle\text{Tr}\left(T_{\mu\nu}\right)\rangle=-\langle T_{tt}\rangle+2\langle T_{x_{i}x_{i}}\rangle=0$. By solving the equations of motion asymptotically near the boundary, one can deduce that $\dot{f}^{(3)}(t)=(3/2)\dot{a}_{0}(t)a^{(3)}(t)$.

In fact, for $d=3$, an analytic result for the expectation values valid for small $\epsilon$ can be obtained by expanding as in (\ref{pertexp}). Adapting the results of \cite{Craps:2013iaa} in our coordinate system, the leading solution in $\epsilon$ for the scalar reads
\begin{align} \label{a1}
a_{1}(z,t)=&a_{0}(t-z)+z\dot{a}_{0}(t-z) +\sum_{m=1}^{\infty}\Big[a_{0}(t-z-2mz_{0})\nonumber \\
&+z\dot{a}_{0}(t-z-2mz_{0})-a_{0}(t+z-2mz_{0}) \nonumber \\
&+z\dot{a}_{0}(t+z-2mz_{0})\Big],
\end{align}
where Neumann boundary conditions have been imposed. Expanding (\ref{a1}) to $\mathcal{O}(z^3)$ specifies $a^{(3)}_1$ and this yields the expression $\langle{\cal O}\rangle$. The relation $\dot{f}^{(3)}_{2}(t)=(3/2)\dot{a}_{0}(t)a^{(3)}_{1}(t)$ then yields the equation for $\langle T_{tt}\rangle$.

$\bullet$ In $d=4$, the counterterms are given by
\begin{align}
\mathcal{S}_{C}=&\int_{\partial\mathcal{M}}\text{d}^{4}\mathbf{x}\sqrt{-\gamma}\left(-6-\frac{1}{2}\hat{R}+\frac{1}{4}(\hat{\partial}a)^{2}\right. \nonumber \\
&\left.+2\ln(z)\left(\frac{1}{16}(\hat{\Box}a)^{2}+\frac{1}{48}(\hat{\partial}a)^{4}\right)\right).
\end{align}
Besides these we can also add finite counterterms,
\begin{equation}
\tilde{\mathcal{S}}_{C}=\int_{\partial\mathcal{M}}\text{d}^{4}\mathbf{x}\sqrt{-\gamma}\left(\frac{\alpha}{8}(\hat{\Box}a)^{2}+\frac{\beta}{24}(\hat{\partial}a)^{4}\right),
\end{equation}
with arbitrary constants $\alpha$ and $\beta$. The expectation values are then scheme-dependent
\begin{equation}
\langle{\cal O}\rangle=4a^{(4)}(t)+\frac{(19-24\beta)}{48}(\dot{a}_{0}(t))^{2}\ddot{a}_{0}(t)-\frac{(3-4\alpha)}{16}\ddddot{a}_{0}(t),
\end{equation}
\begin{equation}
\langle T_{tt}\rangle=-\frac{3}{2}f^{(4)}(t)-\frac{(11-24\beta)}{384}(\dot{a}_{0}(t))^{4}-\frac{(1-3\alpha)}{16}(\ddot{a}_{0}(t))^{2},
\end{equation}
\begin{equation}
\langle T_{x_{i}x_{i}}\rangle=-\frac{1}{2}f^{(4)}(t)-\frac{(1-8\beta)}{384}(\dot{a}_{0}(t))^{4}+\frac{\alpha}{16}(\ddot{a}_{0}(t))^{2},
\end{equation}
where $a^{(4)}(t)$ and $f^{(4)}(t)$ are the $z^{4}$-coefficients in the near boundary expansion. The trace of the stress-energy tensor,
\begin{equation}
\langle\text{Tr}\left(T_{\mu\nu}\right)\rangle=-\langle T_{tt}\rangle+3\langle T_{x_{i}x_{i}}\rangle=\frac{1}{16}(\ddot{a}_{0}(t))^{2}+\frac{1}{48}(\dot{a}_{0}(t))^{4},
\end{equation}
is independent of the finite counterterms and indicates the presence of a matter conformal anomaly. It corresponds to the coefficient of $2\ln(z)$ in the counterterms. This result agrees with the more general expressions obtained in \cite{Papadimitriou:2011qb}. By solving the equations of motion asymptotically near the boundary, one can deduce that
\begin{align}
\dot{f}^{(4)}(t)=&\frac{4}{3}\dot{a}_{0}(t)a^{(4)}(t)+\frac{1}{18}(\dot{a}_{0}(t))^{3}\ddot{a}_{0}(t) \nonumber \\
&+\frac{1}{24}\ddot{a}_{0}(t)\dddot{a}_{0}(t)-\frac{1}{48}\dot{a}_{0}(t)\ddddot{a}_{0}(t).
\end{align}

\end{document}